# galsbi: A Python package for the GalSBI galaxy population model

**Silvan Fischbacher**[1], **Beatrice Moser**[1], **Tomasz Kacprzak**[1,2], **Joerg Herbel**[1], **Luca Tortorelli**[1,3], **Uwe Schmitt**[1], **Alexandre Refregier**[1], and **Adam Amara**[1,5]

**1** ETH Zurich, Institute for Particle Physics and Astrophysics, Wolfgang-Pauli-Strasse 27, 8093 Zurich, Switzerland **2** Swiss Data Science Center, Paul Scherrer Institute, Forschungsstrasse 111, 5232 Villigen, Switzerland **3** University Observatory, Faculty of Physics, Ludwig-Maximilian-Universität München, Scheinerstrasse 1, 81679 Munich, Germany **4** ETH Zurich, Scientific IT Services, Binzmühlestrasse 130, 8092 Zürich, Switzerland **5** School of Mathematics and Physics, University of Surrey, Guildford, Surrey, GU2 7XH, UK

## Summary

Large-scale structure surveys measure the shapes and positions of millions of galaxies in order to constrain the cosmological model with high precision. The resulting large data volume poses a challenge for the analysis of the data, from the estimation of photometric redshifts to the calibration of shape measurements. We present GalSBI, a model for the galaxy population, to address these challenges. This phenomenological model is constrained by observational data using simulation-based inference (SBI). The `galsbi` Python package provides an easy interface to generate catalogs of galaxies based on the GalSBI model, including their photometric properties, and to simulate realistic images of these galaxies using the `UFig` package.

## Statement of need

The analysis of large-scale structure surveys can use realistic galaxy catalogs in various applications, such as the measurement of photometric redshifts, the calibration of shape measurements, also under the influence of source blending, or the modelling of complex selection functions. A promising approach to tackle these challenges is forward modeling of the galaxy population. Several such approaches have recently emerged in the literature, differing in their treatment of noise models, galaxy SED modeling, and population sampling strategies. Apart from our efforts with GalSBI, other approaches include SkyPy (Amara et al., 2021), PopSED (Li et al., 2024), and pop-cosmos (Alsing et al., 2024; Thorp et al., 2024).

GalSBI is a parametric galaxy population model constrained by data using SBI (see Cranmer et al., 2020 for a review). Based on one set of model parameters, a galaxy catalog is generated. This catalog is then rendered into a realistic astronomical image using the `UFig` package (Bergé et al., 2013; Fischbacher et al., 2025). The realism of the image relies on two key components: accurate forward modelling of image systematics such as the point spread function (PSF) and the background noise, and a realistic galaxy catalog. For the former, we refer to (Bergé et al., 2013; Fischbacher et al., 2025; Herbel et al., 2017), while the latter is provided by the GalSBI model.

To produce a realistic galaxy catalog, the galaxy population model must be constrained by data. The first version of the galaxy population model, described in Herbel et al. (2017), uses data from the Suprime-Cam instrument on the Subaru Telescope in the COSMOS field to constrain the model. This model was extended by Kacprzak et al. (2020) to measure cosmic shear with the Dark Energy Survey (for more details, see Bruderer et al., 2016, 2018; Chang et al., 2015). Tortorelli et al. (2020) uses the GalSBI framework to measure the B-band galaxy luminosity function using data from the Canada-France-Hawaii Telescope Legacy Survey. In Tortorelli et al. (2021), the model is applied to measure narrow-band galaxy properties of the PAU survey. Fagioli et al. (2018) and Fagioli et al. (2020) use the model to simulate galaxy spectra of the Sloan Digital Sky Survey CMASS sample. Berner et al. (2024) utilizes galaxies sampled from the GalSBI model to produce a realistic spatial distribution of galaxies using a subhalo-abundance matching approach. Further refinements to the model are described in Moser et al. (2024), where they use Hyper Suprime-Cam (HSC) deep fields to constrain the model to high redshift. The first public release of the phenomenological model, incorporating several model extensions is described in Fischbacher et al. (2025). Additionally, Tortorelli et al. (2025) presents a first version of the GalSBI model based on stellar population synthesis.

With the constrained model, we can generate realistic intrinsic galaxy catalogs for various applications. Rendering the catalogs into realistic astronomical images can help to calibrate the shape measurements of galaxies, also under the influence of source blending. Performing source extraction on the simulated images results in realistic measured galaxy catalogs including the redshift distribution. Furthermore, the impact of selection effects can be easily studied by applying the selection function to the catalogs and directly measuring the impact on the observables.

## Features

The main `galsbi` layer allows the user to generate realistic galaxy catalogs based on published GalSBI models as described in Moser et al. (2024) or Fischbacher et al. (2025). With just a few lines of code, the user can generate an intrinsic catalog, simulate astronomical images or run one of the emulators described in Fischbacher et al. (2025) to obtain a measured catalog. We provide the configuration files of these prepared setups in the package to make it easy for the user to get started. However, starting with one of these setups, the user can easily modify the configuration files to adapt the model to their specific needs.

Furthermore, we provide the catalog generator `ucat` as a subpackage of `galsbi`. In `ucat`, the user can define their own galaxy population model using a variety of model choices with different parametrizations. `ucat` is used by the main `galsbi` layer to generate the catalogs and the user can use `ucat` directly to generate catalogs based on their own model. An overview of the different components described above is given in the Table below.

| Component | Core functionality | Details | **`galsbi` connection** |
|---|---|---|---|
| `galsbi.GalSBI` | A convenience layer to load GalSBI models and create intrinsic and measured catalogs based from them. | Provides predefined configurations to run `ucat` and `UFig` plugins. Configurations can be easily customized. | The main interface for running and customizing workflows in the `galsbi` framework. |
| `galsbi.ucat` | A subpackage implementing the phenomenological galaxy population modeling. | Samples intrinsic galaxy properties like magnitudes, sizes, and ellipticities, and provides the `ucat` plugins | A subpackage in `galsbi` that is also called by the main interface. |

| Component | Core functionality | Details | `galsbi` connection |
|---|---|---|---|
| `UFig` | An external package (see Fischbacher et al., 2025) to obtain a measured catalog based from an intrinsic catalog (e.g. generated by GalSBI) | Adds PSF and background to images, can render images and perform source extraction on them or emulate the transfer function from intrinsic to measured catalog. | `UFig` plugins are used in the predefined configuration files of the `galsbi` interface |

Using the model from Fischbacher et al. (2025), sampling a catalog for an HSC deep field simulation in five bands takes about five seconds. This is faster than simulating a single band with `UFig`. However, the runtime depends on the simulation area, depth, and, to a lesser extent, the chosen galaxy population model.

# The GalSBI model overview

In this section, we give a short overview of the GalSBI model. We focus on the constrained model as described in Fischbacher et al. (2025) but the package offers a variety of model choices and parametrizations that are described in the documentation. For interactive versions of the figures, please refer to the corresponding section in the documentation.

## Luminosity functions

The initial galaxy catalog is sampled from two luminosity functions for the red and blue galaxy populations. The luminosity functions are described by a Schechter function with parameters $\phi^*$, $M^*$, and $\alpha$. The two parameters $\phi^*$ and $M^*$ vary as a function of redshift and `galsbi` includes several parametrizations for these functions. Figure 1 shows the blue and red luminosity functions based on the model from Fischbacher et al. (2025) as a function of redshift as well as a simulated image for this specific choice of luminosity functions. The luminosity function determines the number of galaxies in a given area and the absolute luminosity and the redshift of each sampled galaxies.

## Galaxy spectra

In order to obtain an apparent magnitude, each galaxy is assigned a spectrum using a linear combination of the kcorrect templates (Blanton & Roweis, 2007). The coefficients of the templates are drawn from a Dirichlet distribution such that they sum to one. The resulting total spectrum is then normalized to match the absolute magnitude of the galaxy. The apparent magnitude is calculated by applying reddening due to galactic extinction, redshifting the spectrum and integrating it over the filter band.

## Galaxy morphology

The half-light radii of the galaxies are sampled from a log-normal distribution that depends on the absolute magnitude and redshift. Figure 2 shows the half-light radius as a function of redshift and absolute magnitude for the blue and red galaxy populations based on the model from Fischbacher et al. (2025).

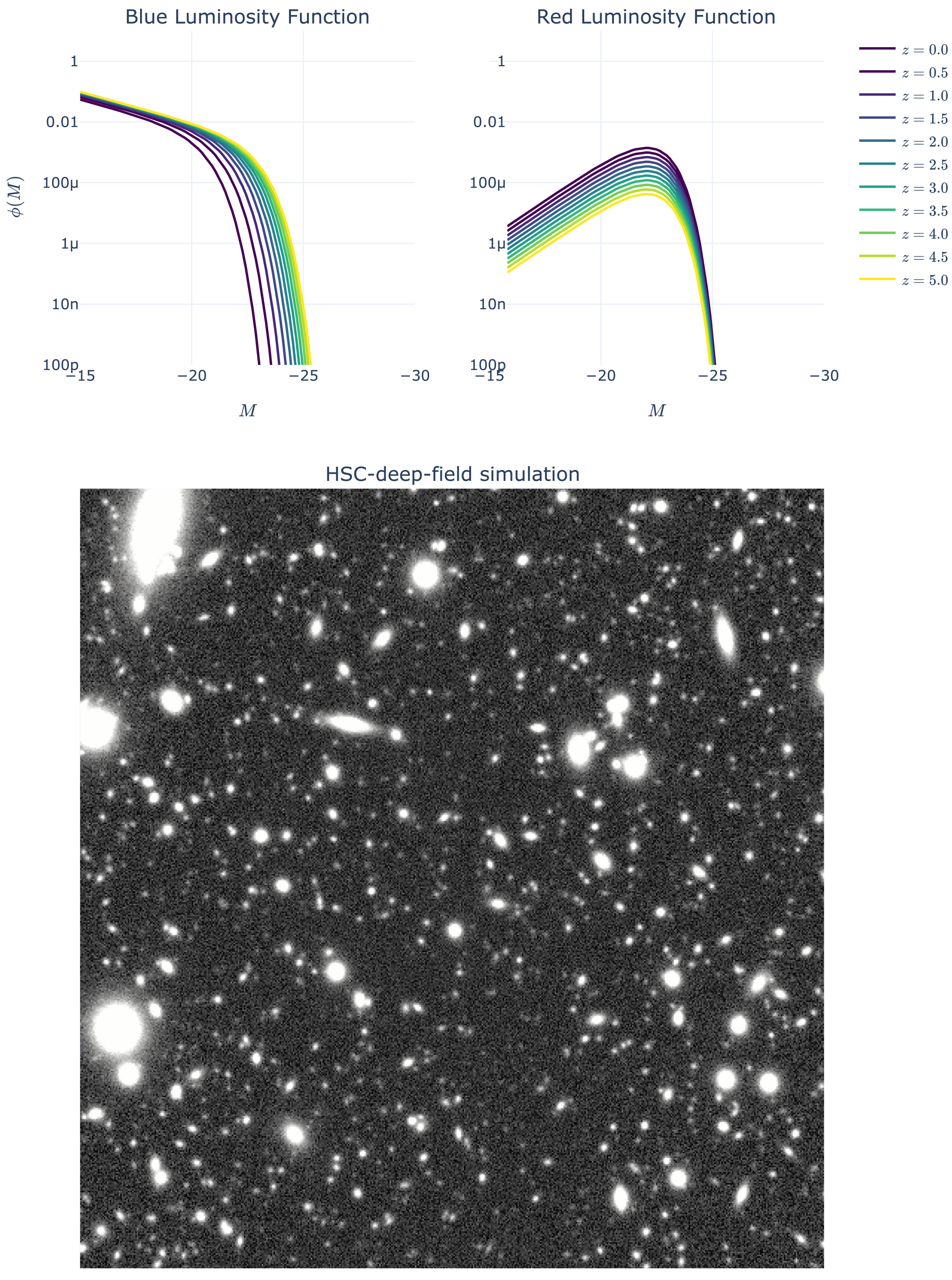

**Figure 1:** Luminosity functions of red and blue galaxies as a function of absolute magnitude $M$. The redshift evolution of the luminosity function is represented by the color gradient, transitioning from low redshift (blue) to high redshift (yellow). The lower panel displays an HSC deep field-like image generated using the above luminosity functions. An interactive version of this plot, including live updates to the image, is available in the documentation.

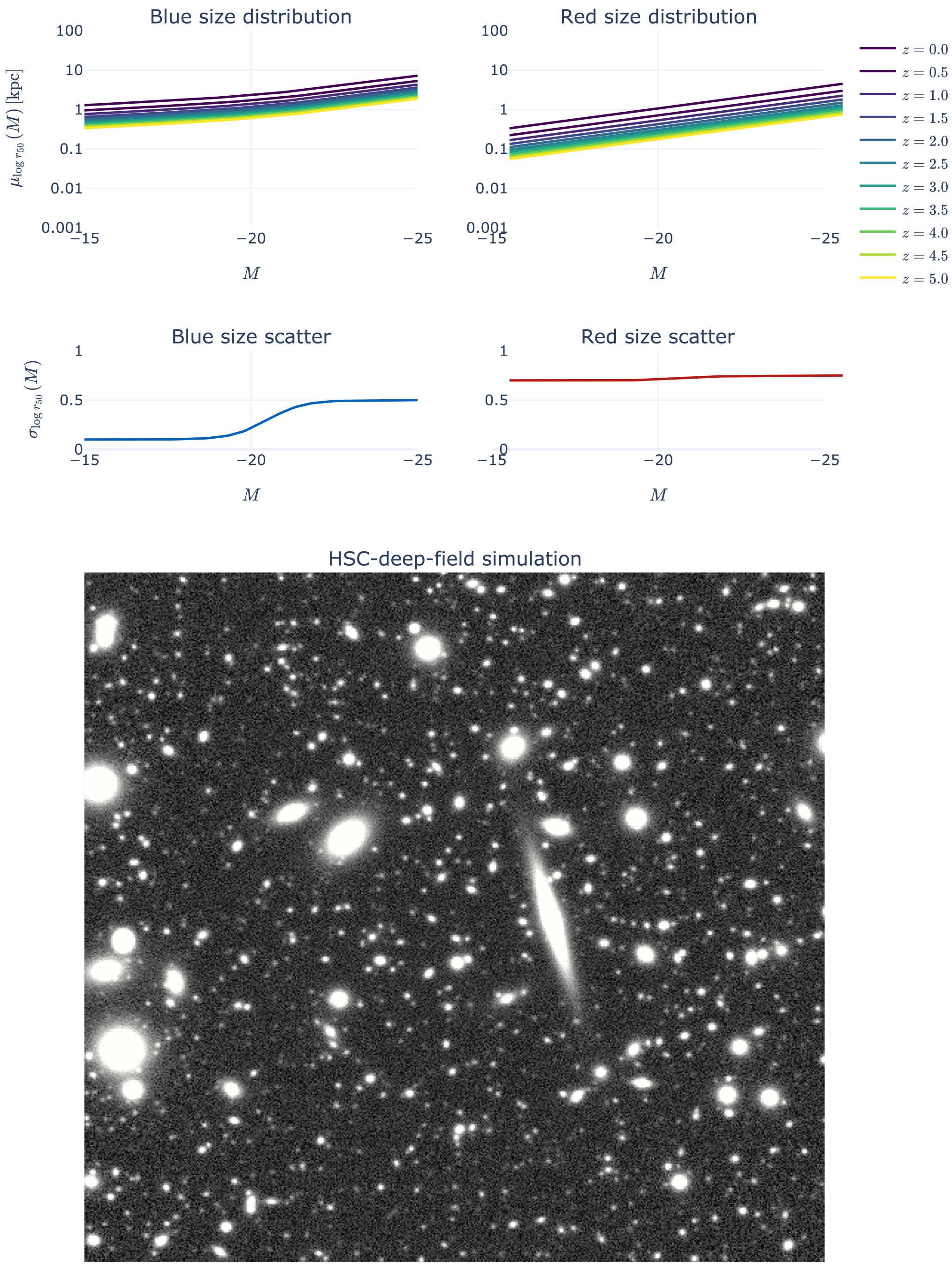

**Figure 2:** Mean and standard deviation of the log-normal size distribution as a function of absolute magnitude $M$ for red and blue galaxies. The redshift evolution of the mean is represented by the color gradient, transitioning from low redshift (blue) to high redshift (yellow). The lower panel displays an HSC deep field-like image generated using the above size model. An interactive version of this plot, including live updates to the image, is available in the documentation.

The ellipticity of the galaxies is defined as a complex number $e = e_1 + ie_2$. UFig requires the two components $e_1$ and $e_2$ to render an image. Depending on the sampling method, the two components are either sampled from Gaussian distributions or the absolute ellipticity $|e| = \sqrt{e^1 + e^2}$ is sampled using different prescriptions and the phase is sampled uniformly.

Finally, each galaxy is assigned a light profile characterized by its Sersic index. In Fischbacher et al. (2025), the Sersic index is sampled from a beta prime distribution.

For more details on the available model choices and parametrizations, please refer to the documentation. A more comprehensive description of the physical motivation of the model can be found in Fischbacher et al. (2025).

## Acknowledgements

This project was supported in part by grant 200021_143906, 200021_169130 and 200021_192243 from the Swiss National Science Foundation.

We acknowledge the use of the following software packages: Astropy (Astropy Collaboration et al., 2013), *healpy* (Zonca et al., 2019), NumPy (Van Der Walt et al., 2011), PyCosmo (Moser et al., 2022; Refregier et al., 2018; Tarsitano et al., 2021), SciPy (Virtanen et al., 2020), and `ufig` (Bergé et al., 2013; Fischbacher et al., 2025). For the plots in this paper and the documentation, we used Matplotlib (Hunter, 2007), Plotly (Plotly Technologies Inc., 2015) and `trianglechain` (Fischbacher et al., 2023; Kacprzak & Fluri, 2022).